# Lattice Dynamics Calculation of the Thermal Expansion of Pure and Ti-Bearing Fused Silica in the Quasi-Harmonic Approximation

James R. Rustad

Corning Incorporated, Corning, NY 14830

The thermal expansion of pure and Ti-bearing fused silica was estimated using free energy minimization lattice dynamics calculations in the quasiharmonic approximation. While the calculations give the expected drop in the coefficient of thermal expansion in going from quartz to fused silica, the overall shape of the coefficient of thermal expansion versus temperature curve for the high-purity and titanium-bearing fused silica is not reproduced by the quasiharmonic approximation. This result is robust and insensitive to system size and annealing density of the glass.

## 1. Introduction

In this report, the coefficient of thermal expansion (CTE) of pure fused silica (HPFS), and fused silica with ~10% added $TiO_2$ is computed from lattice dynamics in the quasiharmonic approximation (QHA) using effective pair potentials taken from the PMMCS force field[1]. The aim is to see whether the addition of $TiO_2$ substantially reduces the predicted CTE of HPFS as observed experimentally, the linear CTE going from around 0.5 ppm/K for HPFS to 0.03 ppm/K in fused silica with ~10 weight percent $TiO_2$. In addition, the PMMCS force field is evaluated for its ability to reproduce the thermal expansion data on crystalline $SiO_2$ (α-quartz) and $TiO_2$ (rutile). The thermal expansion of HPFS is known to be substantially less than α-quartz. It is of interest to see whether a lattice dynamics calculation, with pair potentials fitted to reproduce crystal structures and elastic constants, carried out on glass structures generated from molecular dynamics calculations, would recover this difference.

Thermal expansion is the change in dimension of a body that occurs with a change in temperature. The free energy $F$ of a crystalline system can be written using the following equation[2]:

$$F(V,T) = U_0(V) + \frac{1}{2}\sum_q \sum_i hc\omega_i(\mathbf{q}|V) \qquad (1)$$
$$+ k_b T \sum_q \sum_i \ln\left(1 - e^{-hc\omega_i(\mathbf{q}|V)/k_b T}\right)$$

where $i$ runs over the 3N frequencies, $V$ is volume, $T$ is temperature, $\omega$ is frequency (in cm$^{-1}$)[a], $q$ is a point in the Brillouin zone and the notation "$|V$" (i.e. "evaluated at $V$") is a reminder that the vibrational frequencies are implicit functions of the volume, with larger volumes implying lower vibrational frequencies. $U_0(V)$ is given, in the pair potential model used here, by the following equation:

$$U_0 = \sum_i \sum_{j'} \sum_v \frac{z_i z_j}{r_{ij,v}}$$

$$+ D_{ij}\left[\left\{1 - e^{-a_{ij}(r_{ij,v}-r0_{ij})}\right\}^2 - 1\right] + \frac{C_{ij}}{r_{ij,v}^{12}} \tag{2}$$

where the first sum is over particles $i$, the second is the sum over all particles $j$, and the third sum is over all lattice vectors $v$, with the prime on $j$ indicating that $j \neq i$ if $|v| = 0$. The parameters of this equation, the effective charges $z_i$, as well as $D_{ij}$, $a_{ij}$, $r0_{ij}$, and $C_{ij}$ are given for Si-O, Ti-O, and O-O in Reference 1. Si-Si, Si-Ti, and Ti-Ti interactions have $D_{ij}$, $a_{ij}$, $r0_{ij}$, and $C_{ij}$ equal to zero (i.e. they interact only through the effective charge).

In the QHA, the $\omega_i$ are evaluated in the harmonic approximation. Minimization of the free energy[3], in general, results in a volume expansion of the lattice with respect to the minimization of $U_0(V)$. First, it is immediately clear from Equation 1 that decreasing the $\omega_i$ will decrease the zero-point energy. So, even at 0 K, the lattice will expand to some extent, incurring some penalty in the $U_0(V)$ term to reduce the zero-point energy. At finite temperature, the second term in Eq. 1 makes additional contributions to decrease the free energy as the $\omega_i$ decrease. As the frequency decreases, $\exp(-hc\omega/kT)$ increases (becoming 1 in the limit of $\omega=0$). Because of the $kT \ln(1-\exp(-hc\omega/kT))$ term, the vibrational contribution to the free energy approaches $-\infty$ (i.e. ln[0]) in as $\omega$ approaches zero (see Figure 1 showing the $kT \ln(1-\exp(-hc\omega/kT))$ term as a function of $\omega$ for a temperature of 300K). Since the $\omega_i$ usually become lower with increasing volume (i.e. they decrease with increasing bond length by Badger's rule[4]), increasing the volume of the cell will usually decrease (i.e. make more negative) the vibrational contribution to the free energy. Also note from Figure 1 that the contribution to the free energy is larger (more negative) for a given frequency change low frequency. It is important to remember that it is the natural tendency of the vibrational frequencies to decrease with increasing volume which causes thermal expansion in the QHA, a related but more subtle line of reasoning than the traditional picture of increasing the mean distance between two atoms in vibrating in an anharmonic potential well. Even though the anharmonicity is not explicitly considered in the QHA (the vibrational frequencies are all harmonic) the anharmonicity is manifested as a volume dependence of the vibrational frequencies.

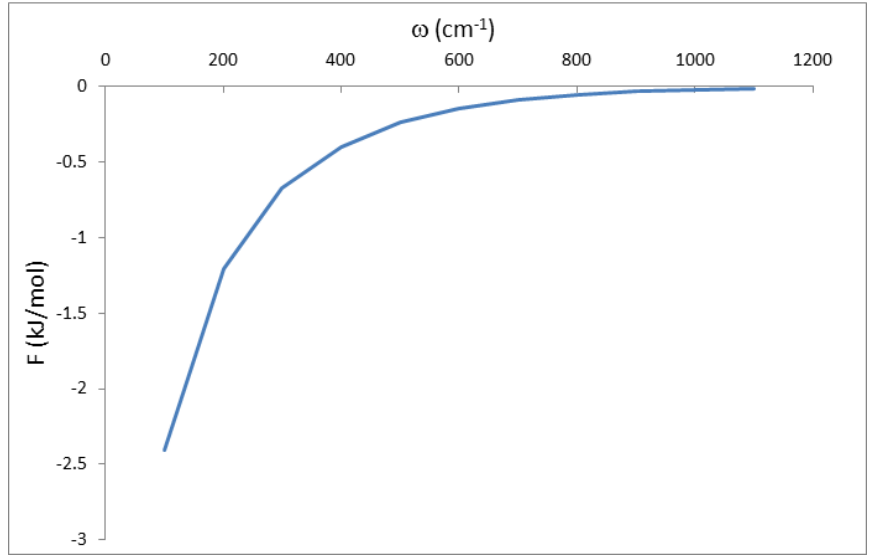

**Figure 1.** *Contribution of oscillator to free energy as a function of frequency ($kT \ln(1-\exp(-hc\omega/kT))$)*

---

[a] In Equation (1) it is convenient to express $c$ in cm/s, $\omega$ in cm$^{-1}$; note that $\omega$ defined as such is not the "angular" frequency, the factor of $2\pi$ which is normally implicit in the definition of angular frequency $\omega=2\pi f$ is not present, and hence Planck's constant is written as $h$ and not $\hbar$.

A couple of remarks are in order concerning the definition of the thermal expansion coefficient since there are two possible sources of confusion. First, one may define the expansion coefficient as either volumetric ($\Delta V/V/\Delta T$) or as linear ($\Delta L/L/\Delta T$), and it is important to specify the convention being used. In this report the linear expansion coefficient is always used and denoted "CTE". Furthermore, since the values are so small, they are normally divided by some small number, which may either be $10^{-6}$ or $10^{-7}$ depending on the convention used. Here, $10^{-6}$ is always used, so that the CTE is reported as ppm/K (parts per million per Kelvin).

Free energy minimization capability is implemented in the GULP code by Gale and co-workers[5]. The zero static internal stress approximation (ZSISA)[6] was employed as has been suggested for silica polymorphs[7].

## 2. Thermal Expansion of $\alpha$-Quartz and Rutile

Before trying to predict the CTE of ULE glass, the PMMCS potentials+QHA combination is benchmarked against the CTE of $\alpha$-quartz and rutile $TiO_2$. This is a reasonable test of the potential functions and the efficacy of the QHA. First, some remarks on a potentially confusing aspect of the PMMCS model. It is stated in Ref. 1 that free energy minimization is used to fit the potential parameters. In trying to reproduce the results of Ref. 1, however, it is found that the parameters in Table 2 of Ref. 1 only yield the structures and elastic constants reported there with straightforward energy minimization (i.e. minimization of *only* $U_0$ in Equation 1). In other words, despite the claims in the paper, free energy minimization was apparently *not* used to fit the potentials. Since the potential parameters reproduce the structures when used in simple energy minimization, the lattice parameters and molar volumes of the crystal are systematically overestimated in free energy minimization calculations both at 0 K and at finite temperature for the reasons just discussed[b]. See, for example, Figure 2 showing the "c" axis of $\alpha$-quartz computed as a function of temperature for the PMMCS force field in the QHA compared with high-temperature x-ray diffraction measurements[8].

---

[b] In many contexts, such as classical molecular dynamics calculations where phonon contributions are not included, it may be more convenient to have parameters fitted to give good structural results without the phonon contribution.

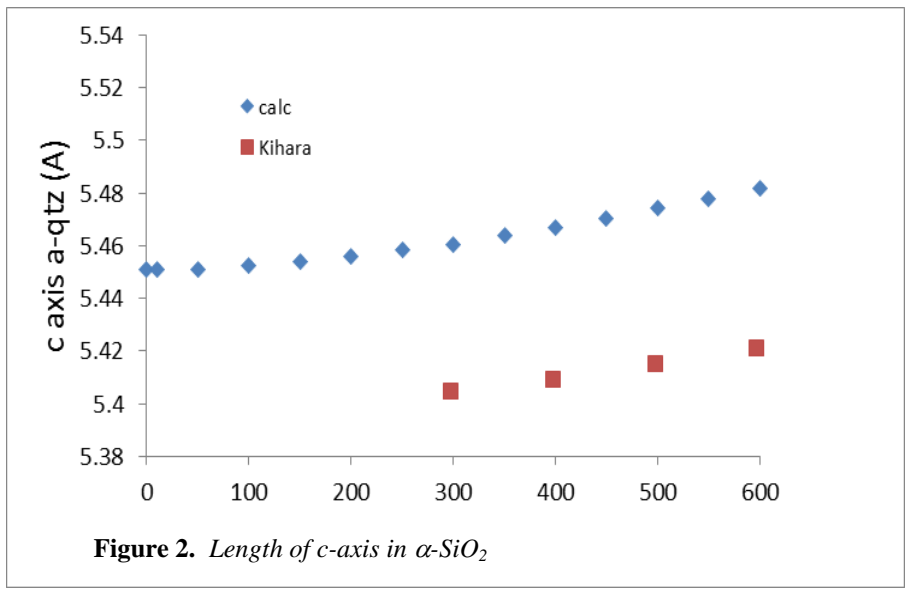

**Figure 2.** *Length of c-axis in α-SiO$_2$*

This overestimation is not too severe, less than, for example, the typical systematic overestimation of the lattice parameters and volume that one gets using the generalized gradient approximation in quantum mechanical treatments.

Predicted CTE as a function of temperature are shown for α-quartz and rutile in Figures 3 and 4 respectively. The data for α-quartz are derived from the x-ray studies[8], and mechanical

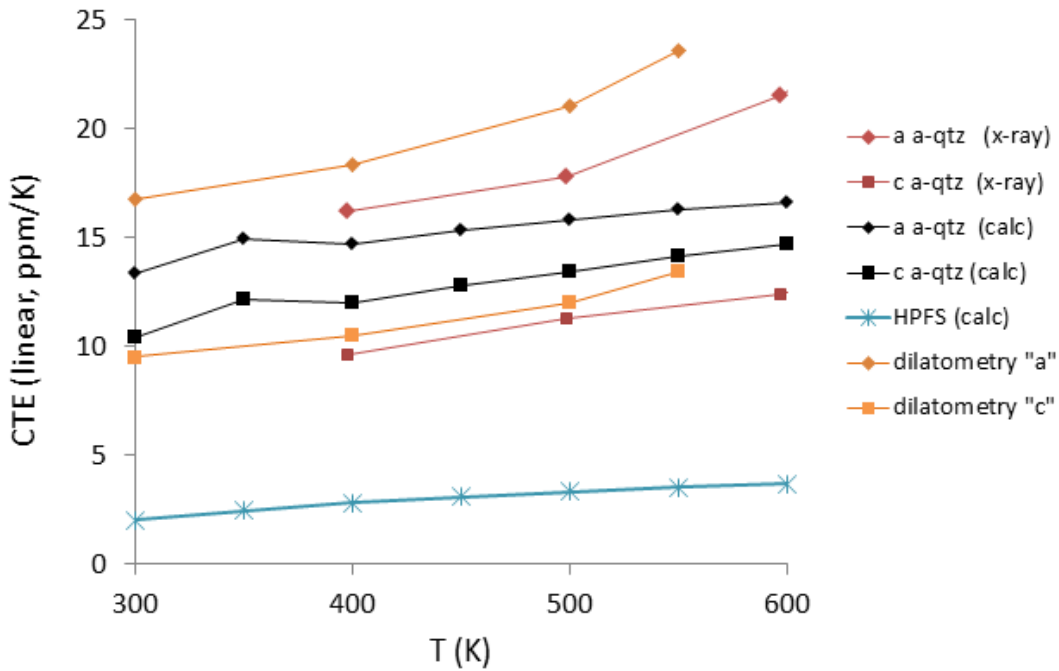

**Figure 3**. *Thermal expansion of α-quartz and HPFS. Measured values taken from Ref. 8 (x-ray) and 9 (dilatometry) . Results for HPFS (discussed below) shown for comparison*

measurements[9] the data for rutile are taken from interferometry studies[10], and x-ray diffraction measurements as a function of temperature[11].

The calculations correctly predict that for α-quartz, the *c* axis has a lower expansion coefficient that the *a* axis, and for rutile, that the *a* axis has a lower expansion coefficient than the *c* axis. It is not possible, at present, to separate possible errors in the QHA from the errors in the PMMCS force field.

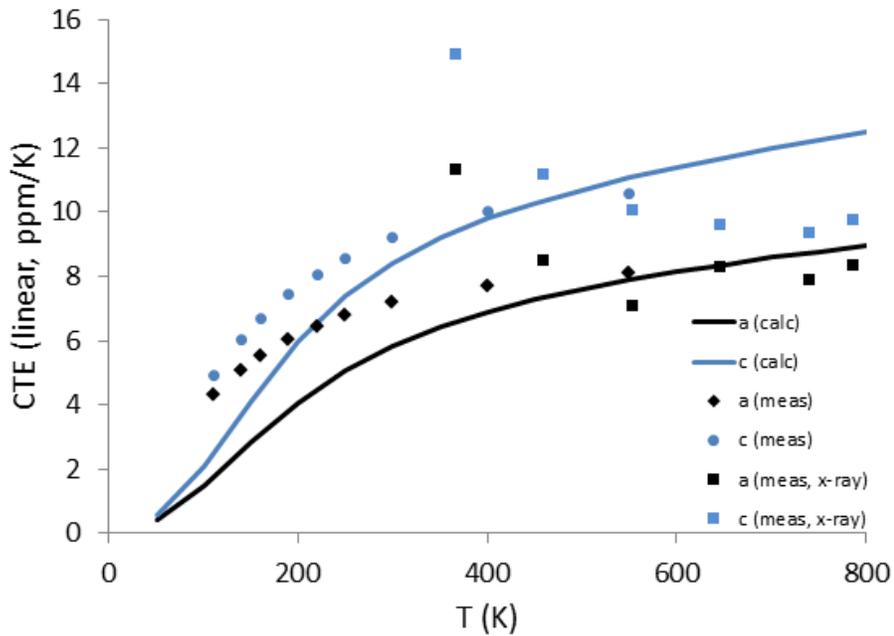

**Figure 4.** *Thermal expansion of rutile $TiO_2$. Measured values taken from Ref. 10 (dilatometry) and Ref. 11 (x-ray measurements).*

## 3. Thermal Expansion of HPFS and Ti-bearing FS

Calculating the CTE of HPFS is a much more difficult problem than the calculation of CTE for the crystalline solids because the atomic coordinates are, of course, not known. To do the calculation, coordinates for the amorphous material must be generated. Here, this is done by simulated annealing in a molecular dynamics simulation. Structures for HPFS and Ti-FS (with

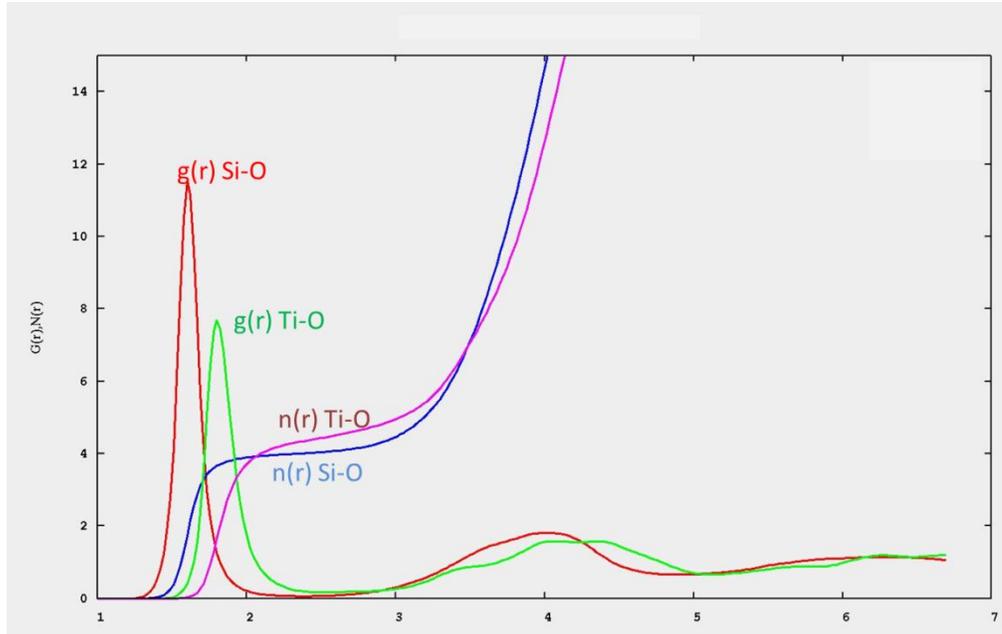

**Figure 5**. *Si-O and Ti-O radial distributions from molecular dynamics generation of amorphous structures*

~10 weight percent Ti) were generated from molecular dynamics calculations using the molecular dynamics code developed by JR Rustad, E Wasserman, JW Halley and A Rahman[12]. The β-cristobalite lattice was heated to 6000K and cooled geometrically (i.e. the temperature is multiplied by a constant factor every timestep) to 300 K over 20 ns. The cell size (13.314 Å) was adjusted to a give a density corresponding to zero pressure at the end of the simulation. A Nose-Hoover thermostat was employed with a mass parameter of 500 amu Å$^2$ .

The glass transition temperature was estimated simply by watching a movie of the dynamics and estimating "by eye" when the silicon atoms ceased to diffuse. Repeated trials give estimates close to 1950 K, as compared to the experimental value of ~1475 K[13]. The overestimation of the glass transition temperature is unsurprising, given the extreme cooling rates associated with the simulated annealing. More sophisticated ways of estimating the glass transition temperature are unlikely to change the estimated value by more than ± 25 K, which is small relative to the disagreement with experiment.

The Si-O and Ti-O radial distribution functions, taken throughout the cooling cycle are given in Figure 5. The Si-O distribution integrates to 4 oxygen atoms under the first peak, the Ti-O distribution integrates to slightly more than four, but this results mainly from the large amount of time spent at high temperature. Below the glass transition temperature, the coordination of Ti is definitively four.

To test the effect of system size and density on the calculated CTE the above procedure was applied to (1) a system of doubled cell size (26.628 Å) and (2) a system annealed at a constant volume with a cell length of 14.077 Å (appropriate to a density of 2.2 g/cm$^3$). Assessment of density is important because the zero-pressure density of the HPFS generated by the PMMCS force field is significantly higher than the measured density of 2.2 g/cm$^3$. After annealing, the final configuration of each run was used as input for a GULP calculation in which the cell parameters and atomic coordinates were geometry-optimized (with no phonon contribution). These configurations were then used in free energy minimization calculations in which the temperature was progressively ramped up from 0K in 50 degree increments.

The results for HPFS are given in Figure 3 where they are compared with data on α-quartz. While the CTE at 300 K is overestimated (2 ppm/K calculated versus 0.5 ppm/K measured) the strongly lowered CTE of the HPFS relative to α-quartz is well reproduced in the calculations. The effects of doubling the system size and decreasing the annealing density are shown in Figure 6. It should be remarked here that only the annealing was done at a density of 2.2 g/cm$^3$ ; the system compacts on straight energy minimization prior to running the free energy minimization. The compacted density is about 2.4 g/cm$^3$. It can be concluded that neither the system size nor the annealing density has a significant effect on the calculated CTE, when compared to the level of disagreement between the calculated (2 ppm/K) and experimental (0.5 ppm/K) values of CTE at 300K, moreover, the temperature dependence of the CTE of the simulated system has a fundamentally different shape than the experimental measurements, strongly following a heat-capacity like curve for a material below the Debye temperature.

The annealing density did have a significant effect on the elastic properties. For the dense system, the bulk and shear modulus are calculated at 68.9 GPa and 40.0 GPa, respectively, while for the less dense system, the values are 42.0 and 34.2.

The calculations for the Ti-bearing fused silica are compared with the HPFS in Figure 7. For the small system, $TiO_2$ actually increases the thermal expansion by a small amount. While in the case of Ti-FS there is some difference between the large and small system, it is clear that the addition of Ti has no effect on the overall shape of the CTE vs T curve, which remains very similar to the shape of a heat capacity versus temperature curve.

## 4. Conclusions

A series of calculations of the thermal expansion of high-purity fused silica and titanium-bearing fused silica were carried out to assess how well the quasiharmonic approximation, when used in conjunction with the PMMCS force field[1] can predict the thermal expansion of these glasses. The potential functions were checked against experimental data for crystalline SiO2 and TiO2 systems. They perform reasonably well, but with room for improvement. One task ahead is to test the sensitivity of these results to the particular force field used. How would, for example, other potentials, such as the BKS potential fare when used in the QHA for HPFS? Ultimately, the most reliable method is to use the QHA in conjunction with potential energy surfaces coming from first-principles calculations.[14]

While the large decrease in CTE in going from crystalline quartz to HPFS is recovered, the overall shape of the CTE curve is poorly reproduced, and the model is not accurate enough to be useful as predictor of CTE behavior in amorphous systems. In particular, the overall shape of the CTE curve, which follows the heat capacity versus temperature curve in the QHA calculations, yet is essentially flat in the experimental measurements, is poorly modeled. As it has been previously shown that it is definitely possible to model even negative CTE materials within the QHA[15], it seems likely that the problem lies in the computational procedure used to produce the amorphous material. While it is shown here that it is not sufficient simply to increase the system size, it seems likely that the problem lies in not taking account relaxation phenomena that are completely outside the realm of the simulation techniques used here. In particular, the simulated glasses take no account of the inherent facilitated dynamical heterogeneities of the glass[16] and their possible coupling to thermal expansion. Recent work suggests that a fundamental feature of the glass transition is the coexistence of self-facilitated mobile islands within large expanses of immobile sea of dead atoms. The implications of this view of glass with regard to thermodynamic properties is just now being worked out[17], and it is interesting to think in particular about the possible implications of this model for the flat temperature dependence of CTE. The unambiguous failure of the QHA to predict the CTE may be a particularly cogent example of the simple properties that fail to be predictable with naïve approaches that do not specifically take into account facilitated, highly correlated ,dynamics of atomic motions below the glass transition temperature.

It should be stressed too that any model that seeks to recover the temperature dependence of the CTE of glass should also reproduce the CTE of rutile and quartz, as done here. There is clearly a relationship between CTE and heat capacity[18] and one way to get a flat CTE vs. temperature curve is to simply use a classical model, which yields a constant heat capacity. One can guard against such illusory success by making sure the model also predicts the CTE versus temperature of the crystalline materials rutile and quartz. The challenge at hand is to predict the CTE versus temperature curves of both HPFS/TiFS and the crystalline phases. The claim made here is that this cannot be done entirely within the quasiharmonic approximation.

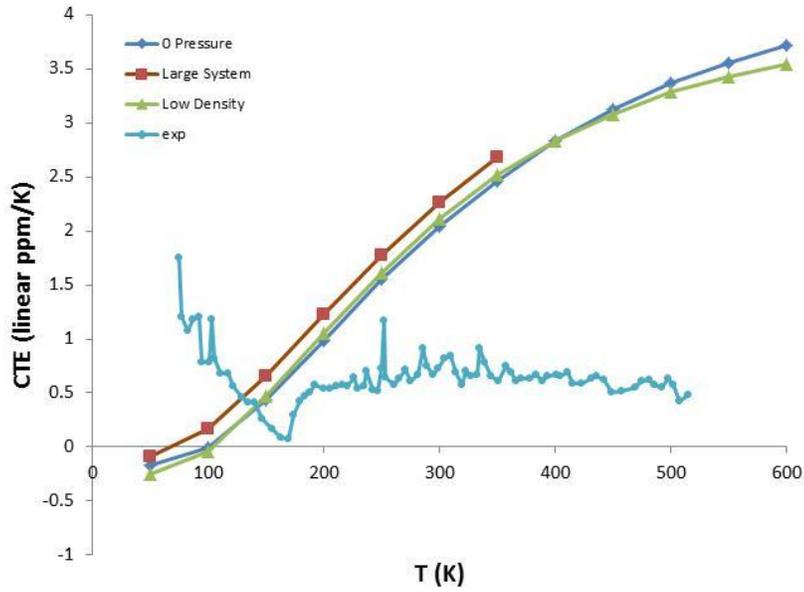

**Figure 6**. *CTE calculated for the large system (double the system size) and low density (2.2 g/cm$^3$) anneal, as compared to the base system. Experimental values are from Corning Incorporated.*

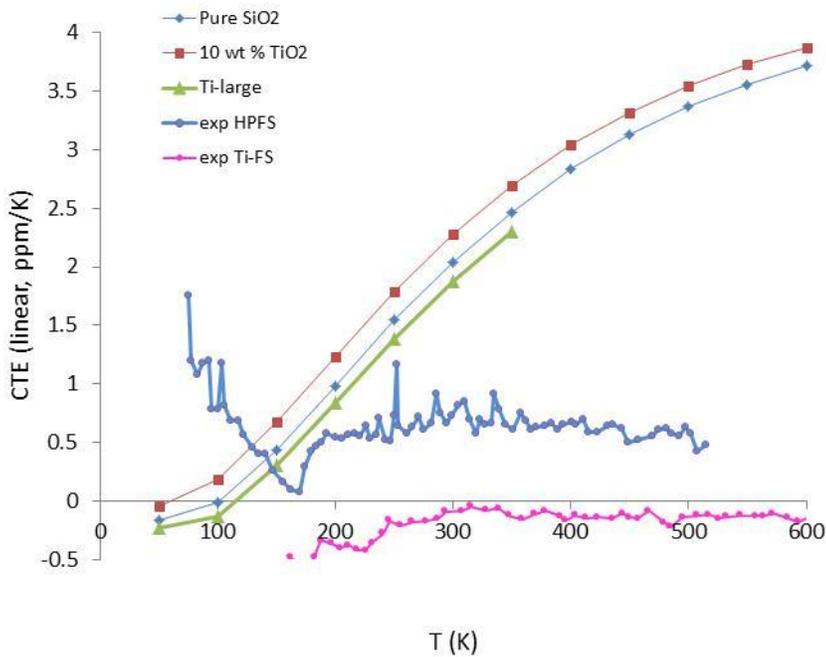

**Figure 7.** *Calculated and experimentally measured CTE for the pure HPFS and Ti-bearing FS. Experimental values are from Corning incorporated.*

Acknowledgements- the author is grateful to Gulcin Tetiker, Ross Stewart, and Aravind Rammohan for discussions, for checking calculations, and for the data presented in Figures 6 and 7.Acknowledgements- the author is grateful to Gulcin Tetiker, Ross Stewart, and Aravind Rammohan for discussions, for checking calculations, and for the data presented in Figures 6 and 7.